# Asymmetric Diffusion

Norman Packard[1,3,4] & Rob Shaw[2,3,4]

[1]ProtoLife S.r.l., via della Libertà 12, Venezia, Italy

[2]University of Modena and Reggio Emilia, Reggio Emilia, Italy

[3]European Center for Living Technology, Venice, Italy

[4]Both authors contributed equally to this article.  Correspondence to Rob Shaw, rob@far-west.net.

Abstract

Diffusion rates through a membrane can be asymmetric, if the diffusing particles are spatially extended and the pores in the membrane have asymmetric structure.  This phenomenon is demonstrated here via a deterministic simulation of a two-species hard-disk gas, and via simulations of two species in Brownian motion, diffusing through a membrane that is permeable to one species but not the other.  In its extreme form, this effect will rapidly seal off flow in one direction through a membrane, while allowing free flow in the other direction.  The system thus relaxes to disequilibrium, with very different densities of the permeable species on each side of the membrane.  A single species of appropriately shaped particles will exhibit the same effect when diffusing through appropriately shaped pores.  We hypothesize that purely geometric effects discussed here may play a role in common biological contexts such as membrane ion channels.

Flows of particles through a membrane may in some circumstances pass more readily in one direction than the other.  This asymmetric diffusion is a known phenomenon in the context of ion transport across membranes (1-3), and in the context of osmosis (4,5). Explanations of these phenomena emphasize binding of particles at intra-pore sites or other electrostatic interactions with the pore surface.  Here we demonstrate that asymmetric diffusion can occur with no binding, through purely geometrical constraints of asymmetric membrane pores on diffusing particles with spatial extent.  Appropriate concentration differences across a membrane may cause complete jamming of the membrane, reducing particle current in one direction to zero.

Consider a chamber containing chemical species or gases divided by a porous membrane. If one of the species is in excess on say the right side, there will be diffusion from right to left through the membrane as the system tends toward equilibrium. The tacit assumption has been that this rate of diffusion would be the same if the species were in excess on the opposite side, that is, the rate of diffusion from right to left is the same as that from left to right. Clearly this must be the case if the membrane is symmetric, i.e., if the pore is symmetric with respect to flipping of the membrane, no matter what the constituent gases or chemicals are. Also, even if the pore structure is asymmetric, if the diffusing particles are very small compared to any pore structure, the point-particle presumption implies that diffusion properties within the pores are identical to those of the bulk. Fick's first law relating the particle flux to the concentration gradient, $\mathbf{J} = -D\nabla C$, must hold everywhere, including throughout the pores, and consequently the diffusion equation $\partial C / \partial t = D\nabla^2 C$ (Fick's second law) holds everywhere, and the rate of particles diffusing in one direction through the pores must be the same as the rate of particles diffusing in the opposite direction, given identical concentration gradients across the membrane. The diffusion equation is linear, with no difference up to a factor between the solutions for a thick gas and a dilute gas, or for that matter, a gas of just one particle.

However, if the membrane pores are asymmetric, and the diffusing particles have spatial extent on the same length scale as the pore structure, so that they interact with each other through collisions between themselves and the structured pores, the diffusion rate toward one side can indeed be different from the rate toward the other, at finite concentration. We will argue that these two diffusion rates are equal in the dilute gas limit, but when concentrations are raised, and particles can interact, the rates can differ, and the symmetry is broken. In the extreme limit the particles can jam the membrane in attempting to pass through in one direction, while being able to flow through freely in the other. This is a strictly geometrical general phenomenon, resulting from the interaction of particle and pore shapes.

We will present two simulation examples of asymmetric diffusion: one is a hard disk gas of two different radii diffusing through a membrane with asymmetrically shaped pores, and the other is a gas of a similar two-species system, acting under Brownian motion, diffusing through a similar membrane. The first has completely deterministic Hamiltonian dynamics, and the second has completely random Brownian dynamics.

**Simulations**

Figure 1 shows typical frames from the hard disk gas simulation. The asymmetric membrane is modeled by a simple beveled pore, wider on one side of the membrane than the other (6-9). There are two species of hard disks, one with a diameter somewhat smaller than the minimum pore size, the other with a diameter somewhat larger, and thus unable to pass through the membrane. In both panels the gas is initially restricted too the right chamber, and then tends toward equilibrium by having the smaller disks move from right to left through the membrane. But in the lower panel, the larger disks are soon swept into the pores, clogging them, and greatly reducing the rate of diffusion of the smaller disks through the membrane. The presence of the smaller disks bombarding the backs of the larger disks makes it highly unlikely that the larger disks will be able to emerge from the pore. In the extreme case the membrane becomes effectively impermeable, rendering the chamber two isolated subsystems.

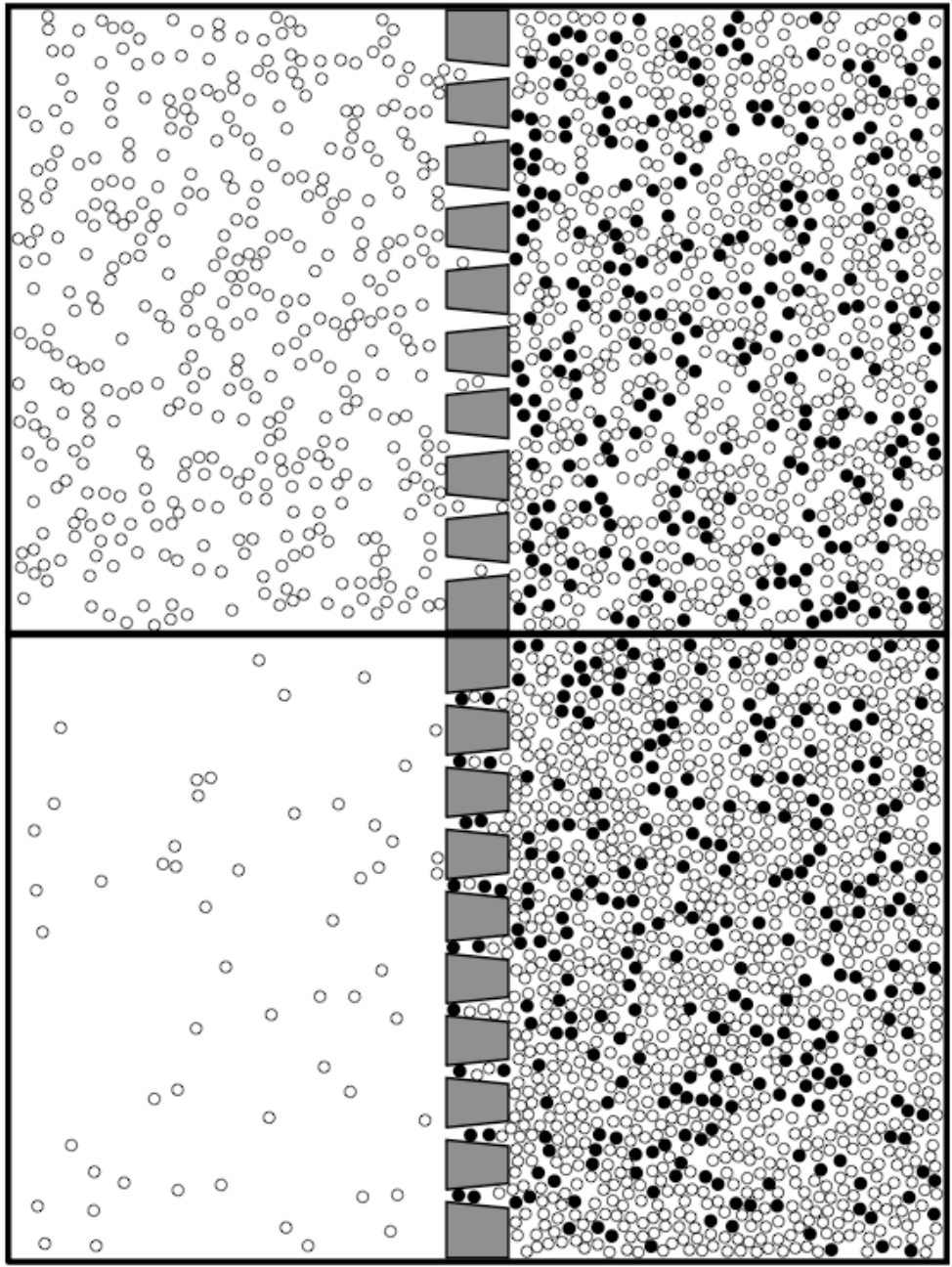

Figure 1. A two-dimensional hard sphere gas, with an asymmetrically shaped membrane, with pores beveled to be narrower on one side than the other. A binary mixture of particles is shown with one species just too large to pass through the pore, and the other just small enough to pass. The left chamber is initially empty. Top and bottom panels show diffusion for the different membrane orientations after identical time periods.

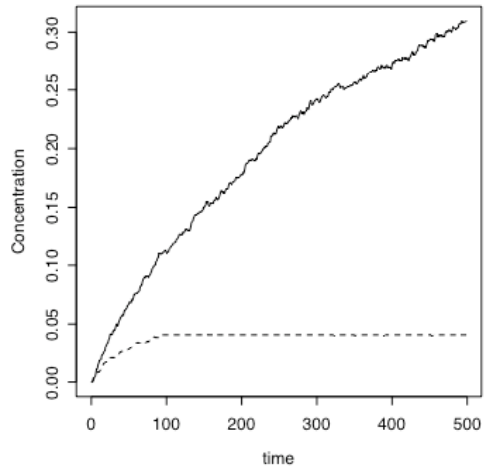

Figure 2 displays a plot of the concentration of the initially empty left chamber as a function of time for the two cases. The lower curve illustrates jamming from the side with shaped (beveled) pores. The diffusion rates across the membrane are the slopes of these curves; in the jamming case the diffusion rate goes to zero.

We see, after running both cases for the same time period, that the flow from the flatter side of the membrane behaves as if there were a well-defined fixed diffusion constant, but the flow from the beveled side drops to near zero, as the membrane seals itself. The clogging behavior of this direction of flow is maintained even at lower concentrations of large disks; just a few percent of large disks can be enough with certain pore geometries. A feature of this simulation is that the more rapidly moving disks preferentially pass through the membrane, resulting in a higher temperature in the side filling up.

We note that this is a purely energy-conserving Hamiltonian system, yet it can spontaneously separate into two subsystems with different temperatures, relaxing to a macroscopic state of disequilibrium. There is a very slow equalization of temperature via collisions at the small ends of the pores, but the separation can be nearly indefinite. Further, when the disk density is lowered and the mean free path of the disks becomes of the same order as the chambers, there occurs the decidedly nonergodic effect of the chamber sizes and shapes affecting diffusion rates through the membrane.

Some insight into the dynamics of this system can be garnered by considering the response to an imposed density variation. The macroscopic state of the system shows a path dependence, as illustrated in Fig. 3. The geometry of the chambers and membrane is the same as in the lower panel of Fig. 1, except that the number of large blocker disks in the right chamber, which we will refer to as the "inside", is reduced from 270 to 200, and there are initially no small disks in either chamber. In the first panel of Fig. 3 the number of small disks in the left chamber, the "outside", is gradually increased at a rate of two disks per time unit, up to 1200 disks. Each disk is added with unit speed, at a random position on the left and moving in a random direction. The number of small disks on the inside roughly follows this increase, as one would expect from a simple diffusion process. The number of small disks on the outside is then reduced at the same slow rate, and again the disks on the inside roughly follow.

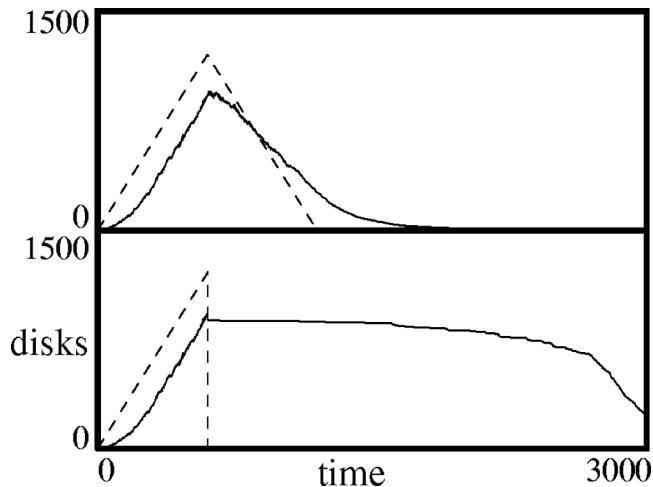

Figure 3. Time-dependent variation of inside concentration of small disks (solid line) in response to variation of outside concentration (dotted line). The first panel shows adiabatic response of inside concentration following the gradual variation of the outside driver; the second panel shows membrane jamming as a consequence of the strong cross-membrane gradient created by the sudden drop of outside concentration.

In the second panel, the same protocol is applied, with a slow increase in the concentration of disks on the outside. When the final number of 1200 outside disks is reached, the outside concentration is again reduced to zero, but this time suddenly, and any small disks leaking out from the inside are removed from the system. In this case, the instantaneous formation of a large concentration gradient across the membrane causes it to jam, and diffusion in the outward direction initially drops to nearly zero, even though the final outside density is identical to that of the first panel.

The diffusion outward does not, however, remain at zero, because the initial concentration of disks on the inside is small enough that there is a significant leakage current. Each small step in the descending portion of the curve in the lower panel corresponds to a single pore opening for a time; a large inside blocker disk is dislodged by a fluctuation, and a "patch clamp" observation of that pore would show a current of small disks flowing outward until the pore was resealed by another large disk entering it. Eventually the concentration of small disks on the inside becomes too small to maintain the membrane jam, and the concentration curve on the inside returns to a simple diffusive decline. Because there are only nine pores in this simulation, the length of the sealed plateau is variable, but the general shape of the knee is repeatable.

The relevance of this jamming effect to observed membrane rectification is we believe evident, and will be examined more carefully in a forthcoming publication. We note in passing that repeated sawtooth driving on the outside will maintain a concentration gradient indefinitely, even with a leaky membrane. Perhaps the simple rectification described here could have played a role in tide-pool scenarios for the origin of life; repeated evaporation and flooding of a primeval tide-pool may have provided the conditions for a sustained ion gradient across the membrane of an early vesicle, enough to drive a proto-metabolism (10).

The embarrassment of riches in the behavior of the hard disk model makes it difficult to study in the limit of low disk density, where we expect to recover classical symmetric diffusion. A Brownian motion simulation we now present has much better behavior in the dilute limit.

Two species of square, one large and one small, move on a lattice, as illustrated in Fig. 4. Once again, the large particles are too large to move all the way through the "membrane", while the small ones can fit (11). No two particles can be at the same lattice location, but at each time step each particle will move, if possible, to a random location among the nearest-neighbor sites that are open. This roughly simulates isothermal Brownian motion. Again, large particles entering from the shaped side can plug the membrane.

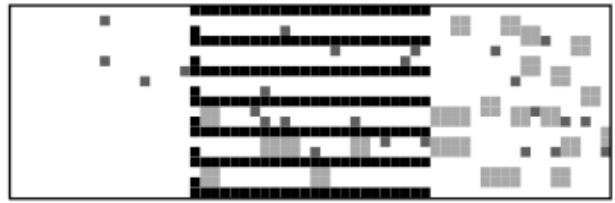

Figure 4. Two species of Brownian particles diffusing through a membrane with asymmetric pores. Jamming of the membrane from its shaped side reduces the diffusion rate.

The two species of Brownian particles show jamming behavior identical to that of the hard-disk gas when particle density is high enough. Figure 5 shows the diffusion rate as a function of initial concentration of objects, for the two orientations of the membrane.

The diffusion rate is defined as the slope of the particle flux curve, averaged over the last 20,000 time steps of a 50,000 step simulation. We see that indeed, in the dilute limit, the diffusion rate is about the same from either side of the membrane. Then there is a Fick's law regime, where the diffusion rate is roughly linear in the concentration, corresponding to a well-defined diffusion constant. Then, as the concentration is further increased, flow in the jamming direction slows, and is finally reduced to zero.

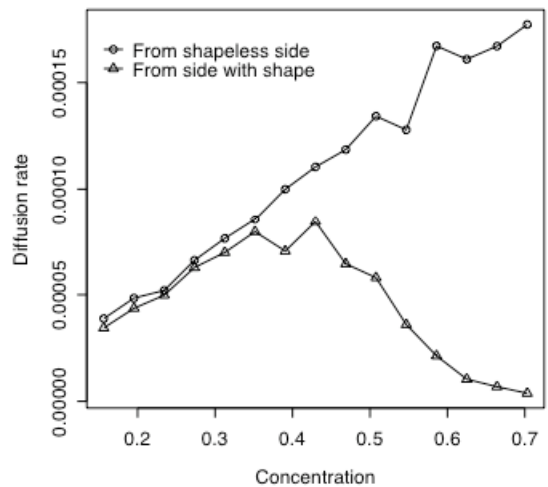

Figure 5. Diffusion rate as a function of concentration. The diffusion constants tend toward the same values in the dilute gas limit, the plugging direction approaches zero as the gas becomes thick.

**Abstract models**

In both of the simulations presented above, there were two species of object, and asymmetric diffusion resulted from a larger object entering a pore first, and impeding the passage of a following smaller object. But the same behavior can readily be produced in passage through an asymmetric membrane of a single species of appropriately shaped object. Instead of two species whose spatial order cannot be reversed within the confines of a pore, we have a single object with two possible orientations within the pore. This situation is illustrated schematically in Fig. 6. When the L-shaped object enters the channel from the left, it must choose between two possible postures, short bar to the left, or short bar to the right. In the former orientation the L-shaped object can continue on past the obstacle, in the latter it cannot, and will jam. A very simple version of single species pore-jamming is given by a domino moving on a lattice (see supplementary material on the web). The difficulty of moving objects among obstacles is well-known in the field of robotics and motion-planning. For example, the "piano mover's problem," moving an object down an arbitrarily shaped corridor, is known to be computationally difficult (PSPACE hard (12)). We have in fact observed some differences in diffusion rates through an asymmetric channel even with a single species of a simple object such as a hard disk or diffusing square; the effect is much weaker than the two-species rectification reported here, and will be reported elsewhere.

However, a simple "domino model" is tractable, and makes clear the possibility of single-species jamming. Imagine a lattice of squares, and a set of dominos sized so that each domino can cover two squares, in either a vertical or horizontal orientation. The allowed transitions are four translations of dominos to neighboring squares, and four rotations bringing a horizontal domino into a vertical position, or vice versa. During rotations, the rule is that one end of the domino must remain on its original square, this generates the four possible rotation transitions. This model is so simple that one can actually draw the full configuration space on a flat sheet of paper. Figure 6 (b) illustrates a domino in a narrow channel, along with the corresponding configuration space, Fig. 6 (c). In the configuration space, vertical domino positions are indicated by circles, and horizontal ones by diamonds. Allowed transitions are indicated by lines between the circles and diamonds. Further, if a domino is occupying some site, no other domino can occupy six of the eight nearest-neighbor sites without overlapping the first domino, as illustrated with the shaded sites in Fig. 6 (c). This exclusion in configuration space represents the domino's spatial embodiment, and is similar in spirit to the exclusion that takes place in the lattice representation of the hard-square gas model (13).

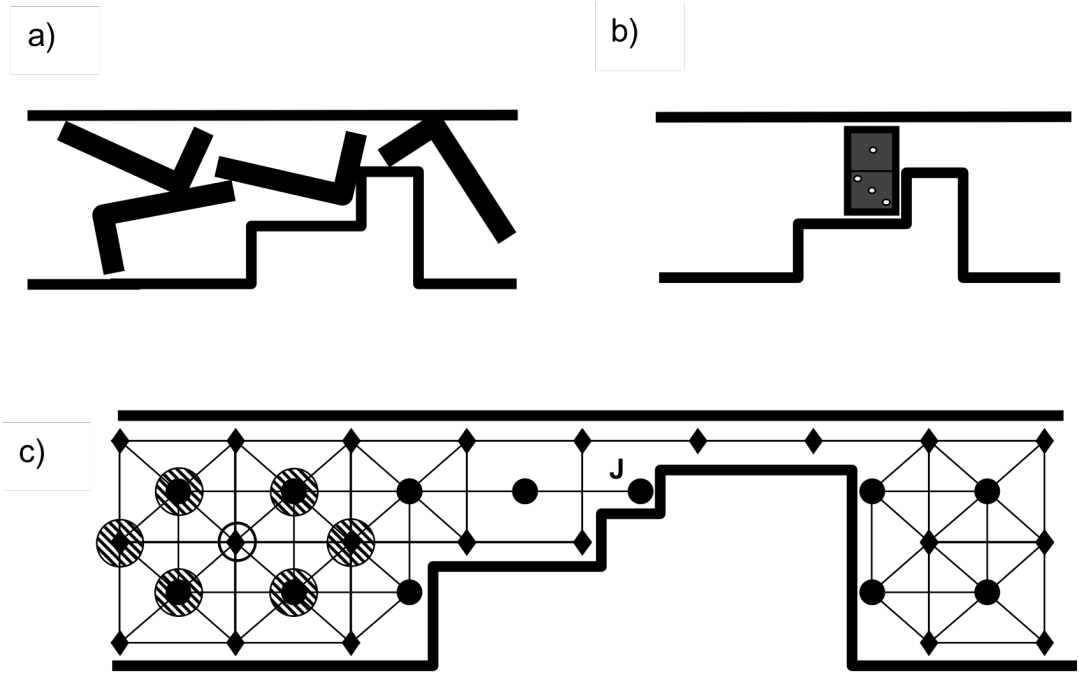

Figure 6. (a) L-shaped objects in a channel with an asymmetric pore. (b) Domino in jammed state. (c) The abstract representation of the domino configuration space, where the jammed state is labeled J. Occupation of the circled site represents a horizontal domino, with shaded circles indicating the configurations excluded by the domino's spatial extent.

Now we can consider the effect of the confining channel geometry at the top of the figure on the configuration space below. The physical boundaries of the channel eliminate the possibility of certain domino positions, this is indicated by the heavy black line boundaries in the configuration space diagram. The domino can only move through the narrow part of the channel in a horizontal position, and so there is only a single transition line connecting the left and right chambers. But the physical boundaries not only delete sites where the domino can reside, they also delete links in the interior of the configuration space. In particular, consider the position of the domino as drawn. In order to pass through to the right, it must rotate to the horizontal position, but it cannot do so in place. It must back up into the chamber on the left, and then perform the rotation. This creates a dead-end alley in the configuration space, with a jammed state at the end, indicated by the letter J. When a domino occupies this state, no other domino may pass through the pore (on the line above) because of the exclusion due to spatial embodiment described above. If another domino comes up behind the first and prevents it from backing up, it will further reduce the probability of the pore becoming unjammed. This is the essence of the phenomena.

The domino model abstracts the motion of continuous shaped objects through a membrane to discrete steps on a class of networks. This motivates constructing a minimal network which can exhibit asymmetric diffusion. Figure 7 exhibits a possibility, consisting of just three nodes. The network is fed objects at one of two possible input nodes, representing the entrances to some pore, at a rate $\alpha$ or $\alpha'$, representing the rate at which the network is fed by some reservoir. The rules are, only one object can be at a node, and objects move randomly along any link that is available. For example, if an object is introduced into the upper-right node from reservoir $\alpha'$, it will with probability 1/2 move to the left node if it is open, or will exit the network to the right, presumably back to the reservoir. The reservoirs are always open. This is a very simple example of a class of kinetic model used some time ago by Hille and Schwartz (14) to describe multi-ion single-file flow through a channel.

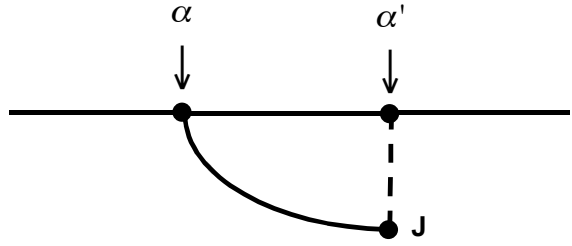

Figure 7. Abstract minimal jamming model, with the jamming state J

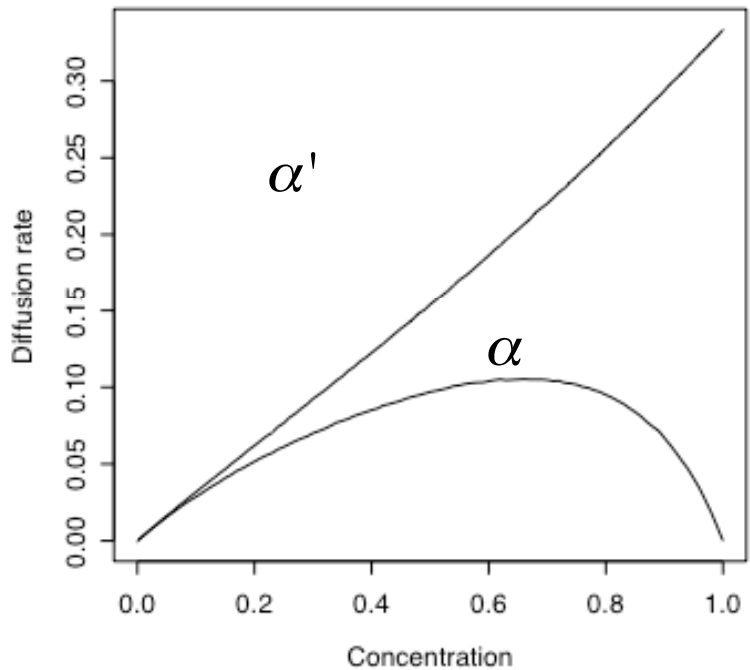

Figure 8. Diffusion rate from abstract model. Compare with Fig. 5.

The lower-right node, marked with a J, is special. An object might enter the left node, and (initially with probability 1/3) have entered this jammed state. It cannot exit to the right, and moreover it prevents objects from entering the upper-right node. The dotted line connecting the two right-hand nodes indicates that they refer to the same spatial location, thus creating the possibility of jamming. The only way to resolve the jam is for the object in the J state to back out the way it came. This may be difficult if the left-hand node is being heavily fed, i.e. the filling rate $\alpha$ is high.

Figure 8 plots the flux across the network as a function of $\alpha$ or $\alpha'$, for example, the number of particles per time step that enter from $\alpha$ and successfully exit the network at the right. One more time we see striking asymmetric diffusion; particles entering at $\alpha'$ generate a flux across the network nearly linear in $\alpha'$, while particles moving in the other direction get jammed. For low $\alpha$, meaning there is is no more than one particle at a time in the network, the flux in both directions is equal, linear in $\alpha$, with slope 1/3. As $\alpha$ goes to one, meaning an attempt to add a particle on the left is made every time step, the flux to the right actually drops to zero. This figure is qualitatively similar to Fig. 5, the diffusion rate across the membrane in the Brownian simulation.

Although our examples and simulations are restricted to two dimensions, or in the case of the network models even lower, three-dimensional analogs are readily constructed.

**Discussion**

Asymmetric diffusion at first examination might seem counterintuitive, and at odds with the Second Law. But there is no contradiction, as this is a nonequilibrium effect, being driven by increased concentration on one side. Nevertheless it is striking, and this phenomenon can occur arbitrarily close to equilibrium. Abstractly, the configuration space is split into two branches when a blocker enters the pore. In one branch, the pore is clogged, and the configuration space terminates in an alley where it is statistically unlikely to emerge. This dead-end feature of a blocker's configuration space was clearly noted by Hille and Schwartz (14), where transitions in the configuration space have rate constants attached as a result of binding effects. In our case there is no energy difference between the two branches when the particle enters the pore, but there is a statistical effect, driven by a difference in chemical potential, and the probability of re-emergence can be made arbitrarily small by making the pore arbitrarily long. These two facts imply that the blocking effect can occur arbitrarily close to equilibrium. This we believe to be relevant to biological situations; the effect can be generated in isothermal conditions, and driven only by a small concentration gradient.

The jamming effect is not directly dependent on temperature. Temperature enters both the hard disk and the Brownian simulations only as a scaling factor of the time; speeding up the objects does not relieve the jam.

Asymmetry has long been noted in movements of ions through channels in membranes, where the asymmetry is often directly described as a form of rectification (1). Although ion channels can be complicated devices, with voltage sensors and internal conformational changes, a particularly simple pore design is the inward-rectifier potassium channel; for a review see Lu (15). Early modeling of this channel explicitly described the outward flow of $K^+$ ions through the channel as being blocked by larger ions which are present on the inside of the cell and cannot pass completely through (14,16-18). However, models typically emphasize electrical effects, in particular intra-pore binding sites, to keep the blockers in place in the pore. Here we suggest that the jamming effect described in this paper may play a key role in maintaining blocking, and creating the effect of rectification.

The detailed atomic structure of a member of the potassium channel family has recently been imaged (19). It shows an asymmetric pore, narrow towards the outside of the membrane, and wider towards the inside, roughly like our model pores represented in Figs. 1 and 4. In addition, there is a chamber toward the inside, which may function to increase the residence time of the blocking ion, thus increasing the rectification efficiency. The inclusion of the jamming effect may help clarify discussions of "multi-ion channels" (14,20). We are not claiming that electrical effects and binding play no role in the dynamics of rectification; they clearly do. But we can claim that any first-order explanation of inward rectification should include the geometrical jamming effects described here. The simple jamming phenomenon has the added attraction of suggesting a clear evolutionary path between a purely mechanical rectification that might occur in a prebiotic vesicle, and the sophisticated membrane ion channels present in living systems today.

One can find in the literature references to "asymmetric permeability", the observed tendency of some molecules to pass more readily through a membrane in one direction than the other (21-23). But the mechanisms are unclear, and it is sometimes assumed that the expenditure of ATP energy is required to produce this effect. With the jamming effect presented here, we see that asymmetric permeability can arise without an external energy source, from strictly geometrical considerations, in isothermal and near-equilibrium situations.

"Facilitated diffusion" (24-27) provides another example of a passive mechanism that selectively affects diffusion rates. In this case, it is believed that protein complexes embedded in a membrane can facilitate the movement of specific shapes of molecule from one side of the membrane to the other.

There is a relation to the "depletion force", observed for example when small spheres of two different sizes are suspended in a fluid, and undergoing Brownian motion (28). Larger spheres tend to spend more time against the walls of the container, where they are entropically favored to reside. Loosely speaking, the larger particles are pinned against the walls by the bombardment from the bulk by the smaller particles. The similarity to the two-species hard-disk simulation is apparent. Though strictly speaking entropy is not defined in a nonequilibrium situation, there have been studies of the depletion force away from equilibrium (29).

There is a considerable literature on "jamming" in bulk materials (30,31). For example, a powder under compression may bear considerable weight in a rigid state, yet flow freely when the compression is relieved. It is well-known in the materials handling literature that pore shapes affect powder flow, and that the same orifice can transfer powder in one direction and jam in the other (31). Crowding of people and vehicles has been modeled, and strong geometrical effects have been noted, e.g. enhanced flow of people leaving a room by placement of a column in front of the exit (33). In each of these cases, the jamming entities are not simple point particles, but are "embodied," in the sense that they have spatial extent. Even the question of static hard-particle packings and their stability appears difficult (34), and the statistical mechanics of such systems has only just begun, e.g. with hard-square lattice gas models and related cellular automaton models (13).

Finally, we must mention work describing directed motion of small particles in shaped pores, under strong external driving. For example, Brownian motion through the pores can be biased in one direction after application of both an applied periodic electric field (35), or applied periodic fluid pressure oscillations (36), even though the driving is itself symmetric. Both produce a potentially useful pumping of particles in one direction, but they are both strongly dissipative. The "nanopore ion pump" (2) is a driven system in this category, operating at a molecular level. The effect we describe is an asymmetry in the unforced diffusion rate, resulting from geometrical and statistical effects, which requires only concentration differences to be seen.

**Conclusion**

We believe asymmetric diffusion to be a rather general phenomenon, occurring in both the interaction of shapes in restrictive geometries, as exemplified by our membrane models, and on networks where there is limited occupancy, with interactions between nodes. Here "diffusion" is broadly defined as motion of objects, with a random component. It is quite natural for large objects to get stuck, and block passage of following objects; all that is required is a certain density and a flow direction, no matter how small the flow. The resulting diffusion phenomenon is perhaps not readily described by a simple PDE such as the diffusion equation. The simplicity of the conditions generating asymmetric diffusion suggest that this effect might occur quite commonly in biological systems, where membranes are ubiquitous and molecules of all shapes and sizes are closely packed (37). Large changes in forward and backward diffusion rates can be effected by small changes in pore geometry, or network weights. We think it more than likely that Nature has taken full advantage of this design possibility.

**Acknowledgements**

We would like to acknowledge Josh Deutsch for a critical reading of the manuscript, as well as Mark Bedau, Jim Bredt, Jim Crutchfield, Doyne Farmer, Kristian Lingren, and John McCaskill for helpful comments. This research was partially funded by PACE (Programmable Artificial Cell Evolution), and by ISCOM (Information Society as a Complex System), European Projects in the EU FP5 and FP 6, IST-FET Complex

Systems Initiative. The authors appreciate hospitality of the Santa Fe Institute and the European Center for Living Technology.